\documentclass{article}
\usepackage{spconf,amsmath,graphicx,hyperref}

\usepackage[T1]{fontenc} 
\usepackage[utf8]{inputenc} 
\usepackage{amsmath,cite,url, amssymb}
\usepackage{graphicx}
\usepackage{pgf-pie}
\usepackage{float}
\usepackage{booktabs}
\usepackage{tabularx}
\usepackage{placeins}
\newcommand{\ra}[1]{\renewcommand{\arraystretch}{#1}}

\usepackage{color}
\usepackage{xcolor}

\newcommand{\dsname}{MulTTiPop}
\newcommand{\dssize}{3.5}
\newcommand{\dssongs}{572}

\title{\dsname{}: A Multitrack Transcription Dataset for Pop Music}
\name{%
  \begin{tabular}{@{}c@{}}
    Nathan Pruyne$^*$\thanks{$^*$ Corresponding author: \href{mailto:npruyne@cmu.edu}{npruyne@cmu.edu}} \qquad Benjamin Stoler \qquad William Chen \\
    Chien-yu Huang \qquad Shinji Watanabe \qquad Chris Donahue
  \end{tabular}%
}
\address{Carnegie Mellon University}
\begin{document}
%
\maketitle
\begin{abstract}
We present \emph{\dsname{}}, a dataset of pop music segments and their associated multitrack MIDI recordings for the evaluation of automatic music transcription models. \dsname{} contains \dssongs{} segments of popular music totaling \dssize{} hours of audio, and contains
songs from diverse genres and decades from the 1930s to 2000s.
To collect this dataset, 
we perform metadata-based matching on song segments from the Lakh MIDI and TheoryTab datasets, manually identify an anchor beat between the audio and MIDI, then use beat tracking on the audio and warp the MIDI to match its tempo and timing.
We evaluate state-of-the-art automatic music transcription models on \dsname{} and find substantial room for improvement, with the best model achieving $29\%$ Onset F1. 
More details and sound examples of \dsname{} are available at \url{https://gclef-cmu.org/multtipop}.
\end{abstract}
%
%

\begin{table*}[!t]
\centering
\ra{1.2}
\begin{tabular}{llllcc}
\toprule
\textbf{Dataset} & \textbf{Music Genre} & \textbf{Annotation Type} & \textbf{Audio Type} & \textbf{Size (Hours)} & \textbf{Total Samples} \\ 
\midrule

MusicNet \cite{thickstun2016learning} & Classical & \textbf{Multitrack MIDI} & \textbf{Commercial} & 34 & 330 \\

MAESTRO \cite{hawthorne2018enabling} & Classical Piano & Single track MIDI & Original & 199 & 1276 \\

Slakh2100 \cite{manilow2019cutting} & \textbf{Popular} & \textbf{Multitrack MIDI} & Synthesized & 145 & 2100 \\

McGill-Billboard \cite{burgoyne2011expert} & \textbf{Popular} & Chords & Features & 60 & 890 \\

Pop909 \cite{wang2020pop909} & \textbf{Popular} & Single track MIDI & None & 60 & 909 \\

TheoryTab \cite{donahue2022melody} & \textbf{Popular} & Melody and Chords & \textbf{Commercial} & 50 & 22000 \\

MIR-ST500 \cite{wang2021preparation} & \textbf{Popular} & Singing Voice & \textbf{Commercial} & 30 & 500 \\

RWC-Pop \cite{goto2002rwc} & \textbf{Popular} & \textbf{Multitrack MIDI} & Original & 7 & 100 \\

\midrule

\dsname{} & \textbf{Popular} & \textbf{Multitrack MIDI} & \textbf{Commercial} & \dssize & \dssongs \\

\bottomrule
\end{tabular}
\caption{Comparison of existing AMT and pop music datasets, and \dsname{}. \dsname{} provides multitrack time-aligned MIDI annotations for commercial pop music. 
}
\label{tab:datasets}
\end{table*}

\section{Introduction}\label{sec:introduction}

In recent years, automatic music transcription (AMT) systems for converting audio to note-level symbolic representations of music have evolved from transcribing solo piano music \cite{hawthorne2017onsets} to targeting performance on
a wide variety of instruments and musical styles \cite{gardner2021mt3}. However, current AMT models do not yet meet the task of multitrack transcription on real-world pop music. Systems such as YourMT3+ \cite{chang2024yourmt3+} 
are only trained on synthetic recordings of pop music, and perform poorly on commercially-produced songs.
In contrast, Sheet Sage \cite{donahue2022melody} performs well on commercial recordings,
but only generates melodies and chord labels, not full multitrack transcriptions. 
While current AMT models require substantial improvement for use on commercial pop music, their systematic evaluation remains difficult due to the lack of a dataset that matches pop audio with ground truth, time-aligned transcriptions. 
Datasets like MAESTRO \cite{hawthorne2018enabling} and MusicNet \cite{thickstun2016learning} provide transcription labels 
for the acoustically narrow domains of solo piano and classical music, respectively. 
 Slakh2100 \cite{manilow2019cutting} contains multitrack MIDI, but only provides synthesized audio, 
 which frequently does not closely match commercial recordings in timbre, especially for vocals. TheoryTab \cite{donahue2022melody} contains original, fully-produced pop music audio, but only provides weakly-aligned melody and chord annotations, 
 not full multitrack transcriptions. 
The closest existing dataset is RWC-Pop \cite{goto2002rwc}, which contains pop recordings and multitrack MIDI. However, it has limited genre diversity and uses songs composed specifically for RWC-Pop, rather than real-world pop music for which AMT systems may be used for. 

\begin{figure}[!t]
    \centering
    \includegraphics[width=1\linewidth]{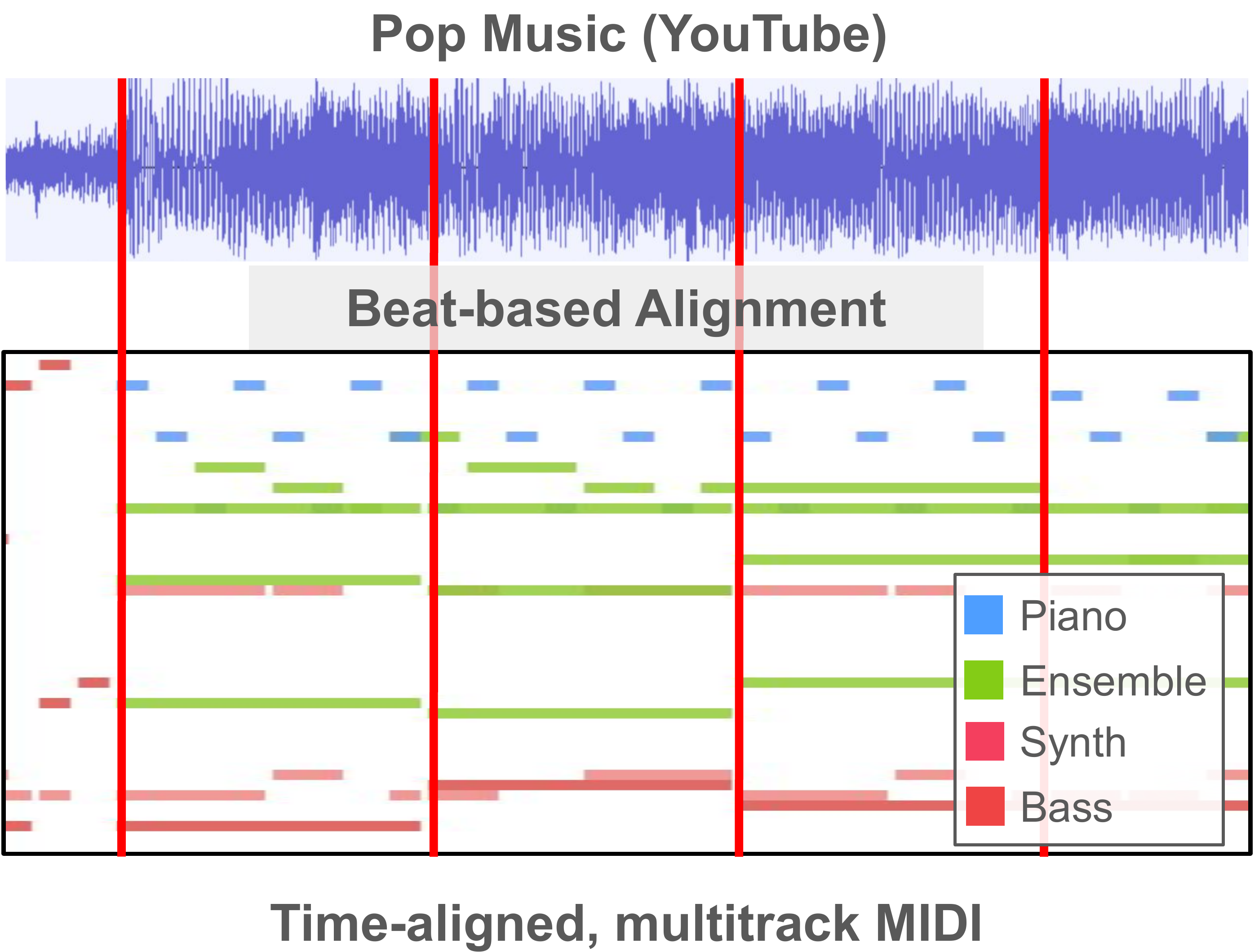}
    \caption{\dsname{} contains segments of YouTube audio beat-aligned with multitrack MIDI transcriptions. 
    }
    \label{fig:fig1}
\end{figure}

To address this gap, \textbf{we introduce \dsname{}, a dataset of multitrack MIDI transcriptions aligned to segments of commercial pop music}. 
We compile multitrack MIDI from the LMD-matched subset of the Lakh MIDI Dataset \cite{raffel2016learning}, and pair MIDI files with audio segments from YouTube by performing metadata matching with the TheoryTab \cite{donahue2022melody} dataset. TheoryTab is sourced from user-provided audio and chord transcriptions, thus indicating segments of audio that users would be interested in transcribing.
\dsname{} has the following key properties:

\begin{itemize}
    \item \textbf{Diverse, Popular Music:} By sourcing audio from TheoryTab, we provide labels for 
    segments of commercial recordings of popular music from the past decades. This enables \dsname{} to be a representative sample of use cases for multitrack AMT.
    \item \textbf{Multitrack MIDI: } For each audio segment, we include a time-aligned MIDI track comprised of many instrument parts present in the original recording. 
    \item \textbf{Commercial Audio: } We list links and timestamps for YouTube videos for all audio segments matching the original recordings.
\end{itemize}

We release \dsname{} as pairs of aligned multitrack MIDI files and associated metadata files with information about the audio.
Specifically, the metadata contains YouTube video IDs and timestamps within the video for the audio that corresponds to the MIDI labels. We make two key recommendations for the use of \dsname{}: 
(1)~researchers should only obtain the relevant segments of the original audio, and 
(2)~researchers should only use this dataset for evaluation, not training.
We compare \dsname{} and related AMT and pop music datasets in Table \ref{tab:datasets}.

\section{Methodology}\label{sec:method}



We outline our approach to aligning audio segments sourced from TheoryTab to multitrack MIDI in the Lakh MIDI Dataset. We first perform metadata matching between audio segments and MIDI files. Then, we time align the MIDI and audio by synchronizing the beat grid of the MIDI with beats detected in the audio. Finally, we identify several candidate anchor beats that align the audio segment to the corresponding portion of the MIDI file, and employ human annotators to select the correct candidate anchor beat that associates with the audio segment.

\begin{figure}[t]
    \centering
    \includegraphics[width=\linewidth]{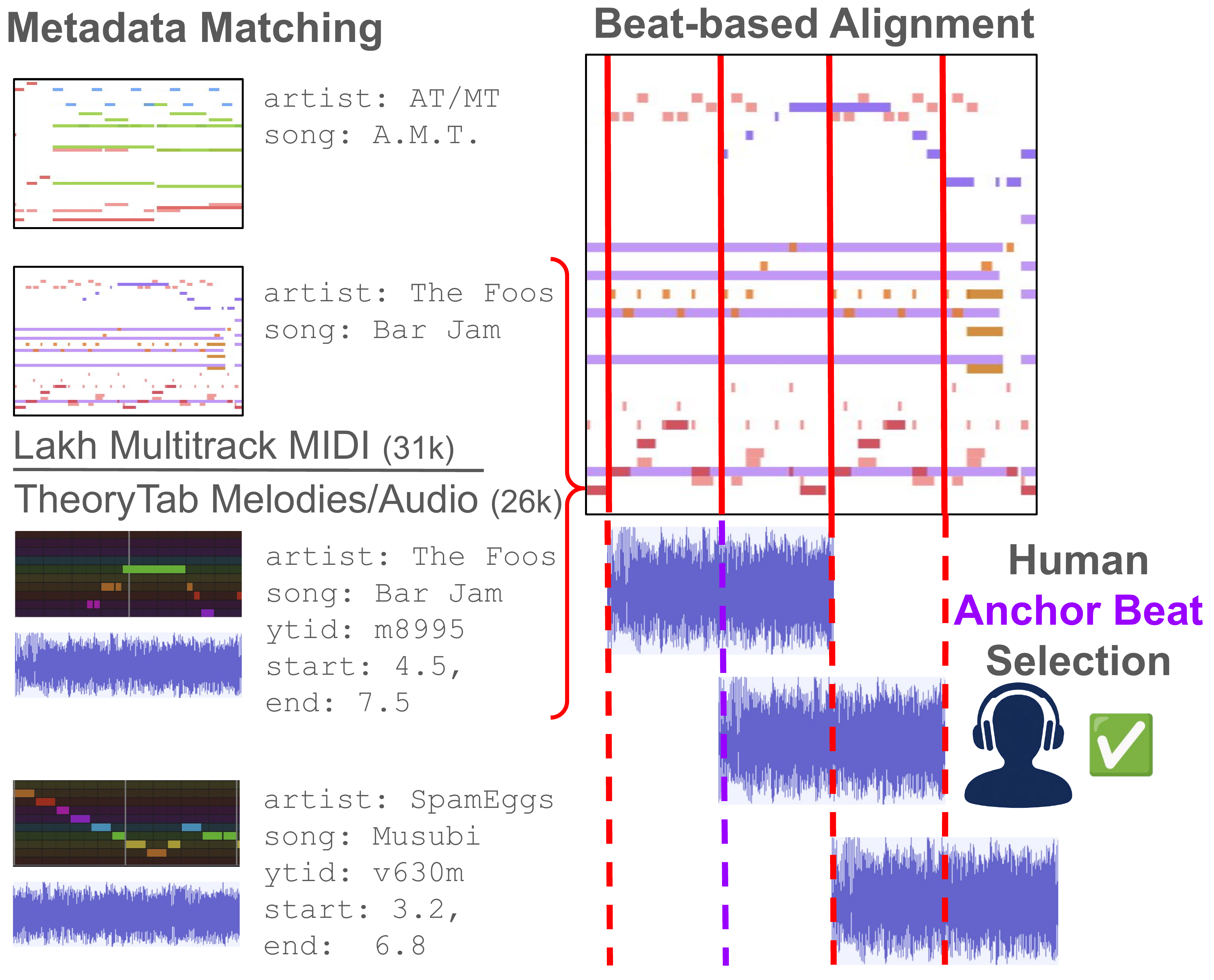}
    \caption{The methodology of creating \dsname{}. We perform metadata matching between the Lakh MIDI Dataset for multitrack MIDI and TheoryTab for YouTube audio information, align MIDI and audio through beat matching, then perform human review to identify an anchor beat in the MIDI that matches the audio segment.
    }
    \label{fig:method}
\end{figure}

\subsection{Metadata-based MIDI/Audio Matching}


We gather audio segments for \dsname{} from TheoryTab \cite{donahue2022melody}, which contains audio through YouTube video IDs with start and end times, user-created melody and chord annotations, and metadata information from the HookTheory platform \footnote{\url{https://www.hooktheory.com/theorytab}}. We begin with 25,947 target segments from TheoryTab, all with valid start and end timestamps, user annotations that are convertible to MIDI, and available audio on YouTube at the time of collection. 
We source multitrack MIDI from the full LMD-matched segment (31,305 songs) of the Lakh MIDI Dataset \cite{raffel2016learning} \footnote{\url{https://colinraffel.com/projects/lmd/}}. This subset has multitrack MIDI transcriptions and metadata from the Million Song Dataset \cite{bertin2011million}, including song title and artist information. We calculate the Levenshtein distance between the TheoryTab and LMD-matched title and artist metadata using RapidFuzz \footnote{\url{https://github.com/rapidfuzz/RapidFuzz}} and filter only to near-identical title and artist fields, yielding 1,164 potential matches between audio segments and multitrack MIDI. 

\subsection{Beat-based Audio-MIDI Time Alignment}

Despite the multitrack MIDI we identify consistently containing the correct musical content, the tempo of the MIDI is slightly different from the audio tempo in many cases, causing poor time alignment between the two. Furthermore, MIDI recordings of pop arrangements are defined with strict beat grids, while actual recordings of songs may have small changes in tempo throughout. Because of this, we perform fine-grained alignment between the multitrack MIDI and audio via beat matching in a similar style to the preprocessing in Sheet Sage \cite{donahue2022melody}.

Given an audio segment $\mathbf{a}$ with YouTube start and end timestamps from the TheoryTab database $[s,e]$ (in seconds) and its corresponding multitrack MIDI file $\mathbf{m}$, we extract their respective beats $\mathbf{b_a} = [b_a^1, \ldots, b_a^N]$ and $\mathbf{b_m} = [b_m^1, \ldots, b_m^M]$, where $|\mathbf{b_a}| = N$ and $|\mathbf{b_m}| = M$. Note that since the multitrack MIDI file contains a full song, while the audio only contains one song segment, in most cases $M \gg N$. 

The MIDI beats $\mathbf{b_m}$ are defined by the MIDI file's beat grid. To obtain $\mathbf{b_a}$, we perform RNN-based beat tracking \cite{bock2011enhanced} on the audio from $[s - 0.5, e + 0.5]$ using the \texttt{madmom} \cite{madmom} package.
We include $0.5$ seconds of padding on either side of $s$ and $e$ to provide proper acoustic context for the beat tracker, and only preserve beats within $[s,e]$ in $\mathbf{b_a}$. We insert beats halfway between each beat of $\mathbf{b_a}$ if the average detected tempo of $\mathbf{a}$ is half of the defined tempo of $\mathbf{m}$, and remove every other beat of $\mathbf{b_a}$ if the average tempo of $\mathbf{a}$ is closer to double that of $\mathbf{m}$.

Given $\mathbf{b_m}$ and $\mathbf{b_a}$, we can generate a time alignment by defining an \emph{anchor beat} $k \in {1, 2, \ldots, M - N}$ 
that links the first audio beat ($b_a^1$) to the $k$-th beat of the MIDI ($b_m^k$).  We can then define an aligned MIDI transcription by performing linear interpolation on the timing of notes within $\mathbf{m}$ with $|\mathbf{b_a}|$ control points, defined by $[(b_a^1, b_m^k), \allowbreak (b_a^2,b_m^{k + 1}), \allowbreak \ldots, \allowbreak (b_a^N, b_m^{k + N - 1})]$ 
. 
We export the aligned MIDI to a new file where each note is warped to the corrected timing determined by our interpolation, creating time-aligned MIDI labels compatible with standard transcription evaluation.

\subsection{Identifying Candidate Anchor Beats}\label{sec:candidates}

In most cases, only one anchor beat $k$ will yield an alignment that matches the original audio content. 
In preliminary testing, we find that automatic algorithms for identifying the anchor beat are unreliable. Thus, we algorithmically select up to 4 candidate alignments, then perform manual human annotation on the resulting candidates. To determine candidate alignments, we calculate a similarity metric between the audio and MIDI, and select the beat with the highest score. 

As a baseline metric, we use chroma similarity and onset correlation between the time-warped MIDI and audio. We equally weight the cosine similarity between the MIDI and audio chroma, which detects strong harmonic matches with timbre invariance, and the Pearson correlation between the onset envelopes, which captures agreement in the rhythmic attack density. The highest score with this alignment is identified as our \texttt{base} candidate anchor beat.

We expand on this methodology and identify three additional candidate anchor beats by incorporating melody annotations and YouTube audio data from the TheoryTab dataset:

\begin{itemize}
    \item \textbf{Melody matching} (\texttt{melody}):  We compute the chroma cosine between TheoryTab's melody annotations and each individual instrument part in the LMD MIDI, and select the highest-scoring instrument part. This value is weighted as 75\% of the overall metric, with the above \texttt{base} metric as the other 25\%.

    \item \textbf{YouTube video timing} (\texttt{yt}):  We use the start time of the TheoryTab YouTube video $s$, either as an absolute time or as a fraction of the video's duration, and identify the corresponding second in the original, non-warped LMD MIDI file. We then only evaluate beats with our \texttt{base} metric where $|b_a^k - s| \leq 10$, assuming that the relative location of the correct starting beat will be similar in both the YouTube video and the full MIDI arrangement.

    \item \textbf{Both} (\texttt{yt\_melody}): The \texttt{melody} metric applied to beats filtered by the \texttt{yt} methodology.

\end{itemize}





\subsection{Human Alignment Annotation}

To select the correct beat alignment out of the generated options, we present all candidates for the anchor beat to a set of annotators. Annotators are given the original TheoryTab audio, a piano roll visualization of the MIDI for each candidate starting beat, a synthesized version of the MIDI for each candidate, and an overlay of the synthesized MIDI and original audio. Details on the annotation interface are provided in Appendix \ref{app:interface}.

We ask annotators to select an option that transcribes the original audio, emphasizing that we expect the synthesized MIDI to be a perfect multitrack transcription. Annotators reject samples without a perfect match, and indicate whether failures are due to the metadata matching (incorrect MIDI content) or alignment process (incorrect anchor beat or tempo). Annotators are given each sample without knowledge of the method used to select the anchor beat, and if multiple methods produce the same anchor beat, the option is only provided once.

We employ 6 annotators, all students at Carnegie Mellon University majoring in Computer Science, Electronic Music, or Music Technology. To ensure data quality, we embed a hidden control set previously labeled by an author for each annotator, and ask annotators with less than 70\% matches with the author to revisit their annotation. The full dataset was split into nine chunks, each containing approximately 120 files. Annotators spent approximately 3-4 hours annotating each chunk, and were compensated with an \$80 Amazon gift card for each chunk they annotated.

\section{Dataset}\label{sec:dataset}

The resulting \dsname{} dataset contains \dssongs{} segments, totaling approximately \dssize{} hours of audio. Segments come from 374 unique songs and 263 unique artists. Each segment is 22 seconds on average, and contains an average of 583 MIDI notes. A preview of \dsname{} including synchronized YouTube video and MIDI playback is available at \url{https://gclef-cmu.org/multtipop}.

We split \dsname{} into two splits: a dev split for development, and a test split for model evaluation. The dev and test splits are generated at approximately a 7:3 ratio, and are stratified across artists such that no artist appears in both splits. The dev split contains 169 segments and 61 minutes of audio, while the test split contains 403 segments and 152 minutes of audio. We do not recommend using \dsname{} for model training (see section \ref{sec:recs}), thus we do not create training and validation splits.

\subsection{Anchor Beat Selection Efficacy}\label{sec:candsuccess}


Overall, annotators selected one of the candidate alignments for 49.1\% of segments. No candidate anchors created satisfactory alignments for the other 50.9\% of segments. 
Of these failure cases, annotators indicated that 12.2\% were likely due to issues with metadata matching. 
In all other failure cases, annotators indicated that the audio segment appeared to be from the correct song, but the alignment was still incorrect due to the incorrect anchor beat being selected, or a tempo mismatch between the audio and MIDI.

Melody matching markedly increases the success rate of our metric in determining the correct starting beat, increasing alignment success from 2.3\% to 31.7\%. This is likely due to melodies being unique to a specific section: while the \texttt{base} method would select sections that would have high harmonic similarity due to chords and other instrument parts that persist throughout the song, melodies often only repeat on the level of song sections, rather than every measure or chord progression.

Using the YouTube start time to reduce the potential anchor beats to a smaller set proved to overall increase beat selection effectiveness, increasing success from 31.7\% with melody matching alone to 40.0\% with melody matching and the use of the YouTube video timing. We note however that in 109 examples, the \texttt{melody} method would succeed while \texttt{yt\_melody} method would fail, indicating that this method's inherit assumption that the YouTube video and MIDI file would share similar structure and timing would sometimes be incorrect. This suggests that the YouTube videos may be music videos or fan videos which could contain different arrangements of a song or additional content.

\begin{table}[t]
    \centering
    \begin{tabular}{lcr} 
        \toprule
        \textbf{Method} & \textbf{Selected Alignments} & \textbf{Selection Rate} \\
        \midrule
        \texttt{base}       & 25  & 2.3\%  \\
        \texttt{yt}         & 58  & 5.3\%  \\
        \texttt{melody}     & 346 & 31.7\% \\
        \texttt{yt\_melody} & 437 & 40.0\% \\
        \bottomrule
    \end{tabular}
    \caption{Number of alignments where an annotator selected an anchor beat identified by a certain method.
    }
    \label{tab:alignmethods}
\end{table}

\subsection{Dataset Diversity}\label{sec:datadiversity}

\dsname{} contains a wide range of songs within the overarching pop genre, ranging from rock to New Wave to Motown. Dsepite some historical outliers from modern "popular music" (e.g. 1930s swing), the dataset predominantly matches our desired characteristics of containing fully-produced tracks that are relevant to real-world AMT use cases. \dsname{}'s component songs have release years ranging from 1939 to 2009, with the greatest number of songs being published in the 2000s compared to other decades. A detailed breakdown of our definitions of genres and more analysis of \dsname{} is available in Appendix \ref{app:stats}.

The diversity of \dsname{} exceeds that of RWC-Pop \cite{goto2002rwc}, despite its reduced size in hours. We compare various aspects of RWC-Pop, the overall RWC Database \cite{goto2003rwc}, and \dsname{} in Table \ref{tab:rwcompare}. \dsname{} contains songs from over double as many genres and artists as the whole of the RWC database for modern genres.


\begin{table}[t]
    \centering
    \begin{tabular}{lccc}
        \toprule
        \textbf{Dataset} & \textbf{Songs} & \textbf{Artists} & \textbf{Genres} \\ 
        \midrule
        \textbf{RWC-Pop} & 100 & 34 & 2 \\ 
        \textbf{RWC (All Modern)} & 250 & 83 & 43 \\ 
        \midrule
        \textbf{\dsname{}} & 374 & 263 & 101 \\
        \bottomrule
    \end{tabular}
    \caption{Comparison of RWC and \dsname{}. For our analysis, RWC (All Modern) contains RWC-Pop, RWC-Jazz, and RWC-Genre. 
    }
    \label{tab:rwcompare}
\end{table}



\begin{table*}[!t]
    \centering
    \begin{tabular}{lcccccc} 
        \toprule
        & \multicolumn{3}{c}{\textbf{Exact}} & \multicolumn{3}{c}{\textbf{Harmonic-Percussion}} \\
        \cmidrule(lr){2-4} \cmidrule(lr){5-7}
        \textbf{Model} & \textbf{Precision} & \textbf{Recall} & \textbf{Onset F1} & \textbf{Precision} & \textbf{Recall} & \textbf{Onset F1} \\ 
        \midrule
        \textbf{MT3 \cite{gardner2021mt3}} & 11.58 & 10.97 & 10.87 & 29.59 & 28.23 & 27.82 \\ 
        \textbf{YourMT3+ \cite{chang2024yourmt3+}} & 12.88 & 11.25 & 11.51 & 32.17 & 28.11 & 28.74 \\ 
        \bottomrule
    \end{tabular}
    \caption{Evaluation of AMT models on \dsname{} using exact instrumentation and harmonic-percussive instrument reduction. These results suggest that MulTTiPop is a challenging transcription benchmark with substantial headroom for improvement.
    }
    \label{tab:mt3}
\end{table*}

\section{AMT Model Evaluation}\label{sec:eval}

We run two high-performing open weights transcription models---MT3 \cite{gardner2021mt3} and YourMT3+ \cite{chang2024yourmt3+}---and report the Onset F1 of both models in Table \ref{tab:mt3}. 
We evaluate these models with the standard 50 ms tolerance used for Onset F1 \cite{bay2009evaluation} and two definitions of instrument spaces: "exact", where instrument labels must match the same MIDI program as defined in the dataset, and "harmonic-percussive" where the instrument space is reduced to harmonic (pitched) and percussive (unpitched) instruments.

Notably, while
YourMT3+'s architectural improvements and expanded training set allow it to outperform MT3 on other transcription datasets, we do not see a similar performance gap on \dsname{}. Both models also underperform on \dsname{} compared to their reported performance on benchmarks such as MusicNet \cite{thickstun2016learning} and GuitarSet \cite{xi2018guitarset}. We hypothesize that this may be due to both models only training in the multitrack pop space using Slakh2100 \cite{manilow2019cutting}, a synthetic dataset. While YourMT3+ does train on MIR-ST500 \cite{wang2021preparation} for vocal transcription of pop music, its relative small size compared to other training datasets and its use of only singing voice transcriptions likely contribute to it not improving performance on multitrack pop transcription. Both models' transcriptions are available to view on our web preview (see Section \ref{sec:dataset}).

\subsection{Qualitative Model Analysis}\label{sec:qualmodel}


While MT3 and YourMT3+ give transcriptions generally matching the tonality and chord progressions of the samples, they regularly fail to produce coherent transcriptions especially for texturally dense music. In some cases, these models are even able to track melody and vocal lines within songs with some degree of accuracy, but do not transcribe consistent harmony around these lines.

MT3, especially, fails to transcribe multiple lines at once, and will  latch onto one or two parts of the song to transcribe at any given point. This leads to disjointed transcriptions, especially since the instrument parts that it focuses on are not consistent across the segment. MT3 will also label certain lines with different instrument programs throughout the segment, and will sometimes drop notes entirely within lines.

While YourMT3+ does maintain more consistency across parts of an audio segment, this comes at the cost of missing instrument parts and a lack of diversity in the chosen instrument program. We note that YourMT3+ shows improvement over MT3 in tracking vocal lines, especially (however, its chosen instrument program of a saxophone for vocal lines is largely inconsistent with ground truth labels, causing this not to be reflected in the "exact" statistics). YourMT3+, however, appears to have less rhythmic nuance than MT3, and will frequently have simpler rhythms in its transcriptions. 



\section{Recommendations for Use}\label{sec:recs}

\dsname{} is strictly designed as an \textbf{evaluation dataset for automatic music transcription models on multitrack pop music}. \dsname{} is not designed for the training of AMT systems or other machine learning models, especially generative models, both because of the small size of the data and because the labels pertain to copyrighted commercial audio.

We further acknowledge that while \dsname{} is intended to give a representative sample of music that users may wish to transcribe, there is a strong Western bias in our data preparation and presentation. The transcriptions we provide are based on Western harmony and notation. Because our segments of interest are derived from HookTheory, an American company with a primarily Western user base, the songs included in \dsname{} are from predominantly American artists and represent the transcription interests of a Western, not global audience. Because of these limitations, we recommend that model results on \dsname{} should be considered largely a reflection of their performance on music in the Western popular music canon, and not representative of performance on music as a whole.





\section{Acknowledgements}

This work was supported by funding from Sony AI, and we thank our Sony AI collaborators for helpful conversations.
We are also grateful for the contributions of our data annotators at Carnegie Mellon University: Alex Cheng, Woody Li, Arisa Okamura, Sean Xue, Lynn Ye, and Ryan Zhang.
Finally, we thank Song-Ze Yu for flagging issues with our initial over-reported initial evaluation numbers due to a pitch tolerance error, and helping verify the fixed numbers present in this version.

\bibliographystyle{IEEEbib}
\bibliography{ISMIRtemplate}

\clearpage
\appendix

\section{Annotation Interface}\label{app:interface}

\begin{figure}[!h]
    \centering
    \includegraphics[width=1\linewidth]{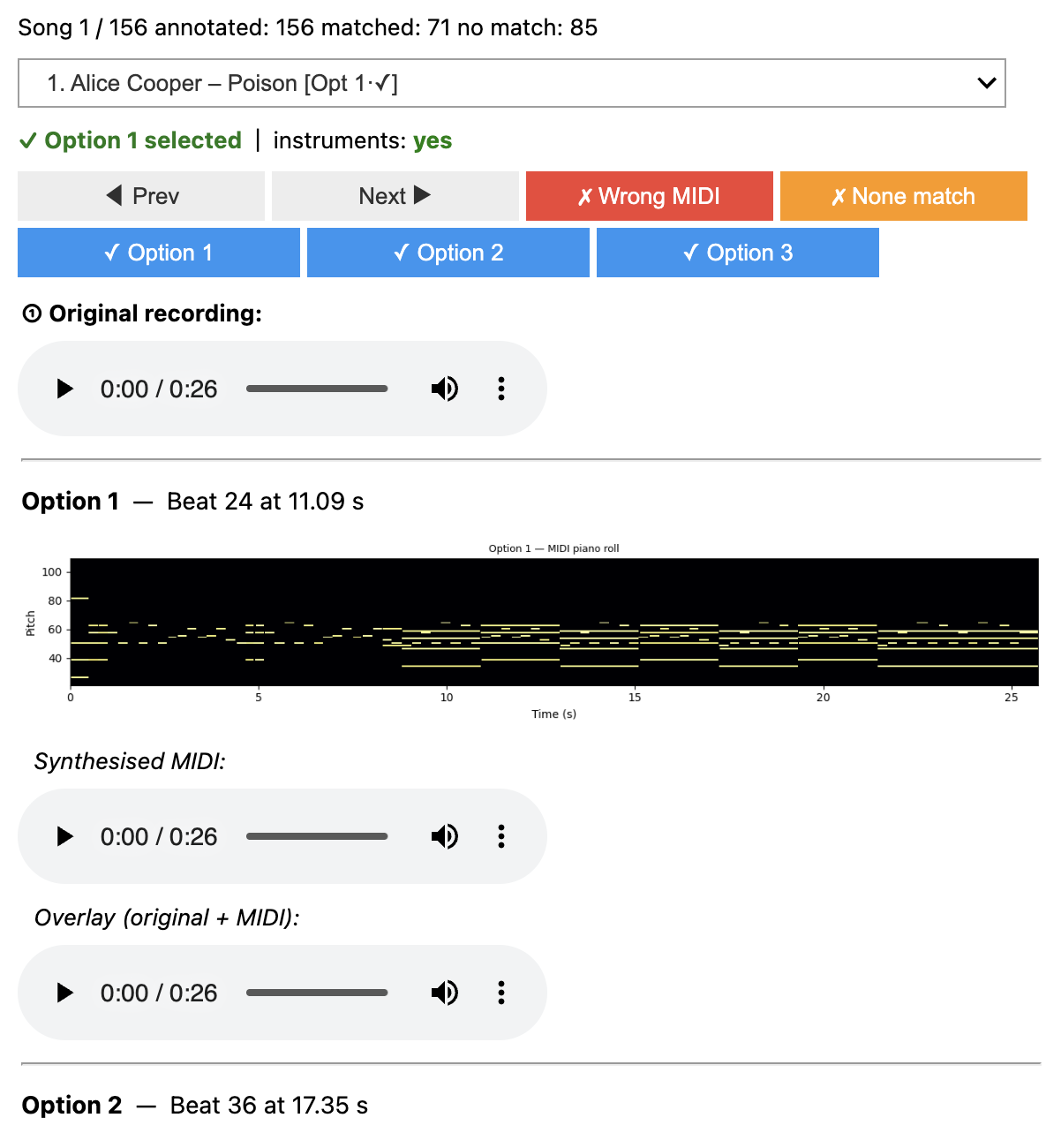}
    \caption{Interface for annotators to select anchor beats among candidate options 
    }
    \label{fig:reviewinter}
\end{figure}

Data annotators are provided a Jupyter notebook (hosted locally or on Google Colab) to perform manual annotation. We present the interface given to annotators in Figure \ref{fig:reviewinter}.

\FloatBarrier

\section{Detailed Dataset Statistics}\label{app:stats}


\subsection{Release Years}


We use Claude Sonnet 4.6 to look up release years for all component songs in \dsname{}. The most common release decade in \dsname{} is the 2000s, with the 1990s, 1980s, and 1970s also having strong representation. The release date of LMD-Matched limits the recency of annotations, so no songs from the 2010s onwards are present. Annotations in pre-1970s decades are increasingly sparse. Full release year information is shown in Table \ref{tab:decades}.

\begin{table}[h]
    \centering
    \begin{tabular}{lr}
        \toprule
        \textbf{Decade} & \textbf{Segments} \\
        \midrule
        1930s       & 3   \\
        1950s       & 1   \\
        1960s       & 28  \\
        1970s       & 108 \\
        1980s       & 128 \\
        1990s       & 136 \\
        2000s       & 168 \\
        \bottomrule
    \end{tabular}
    \caption{Release decade of songs in \dsname{}
    }
    \label{tab:decades}
\end{table}


\subsection{Genre Methodology}

\begin{figure*}
    \centering
    \includegraphics[width=\linewidth]{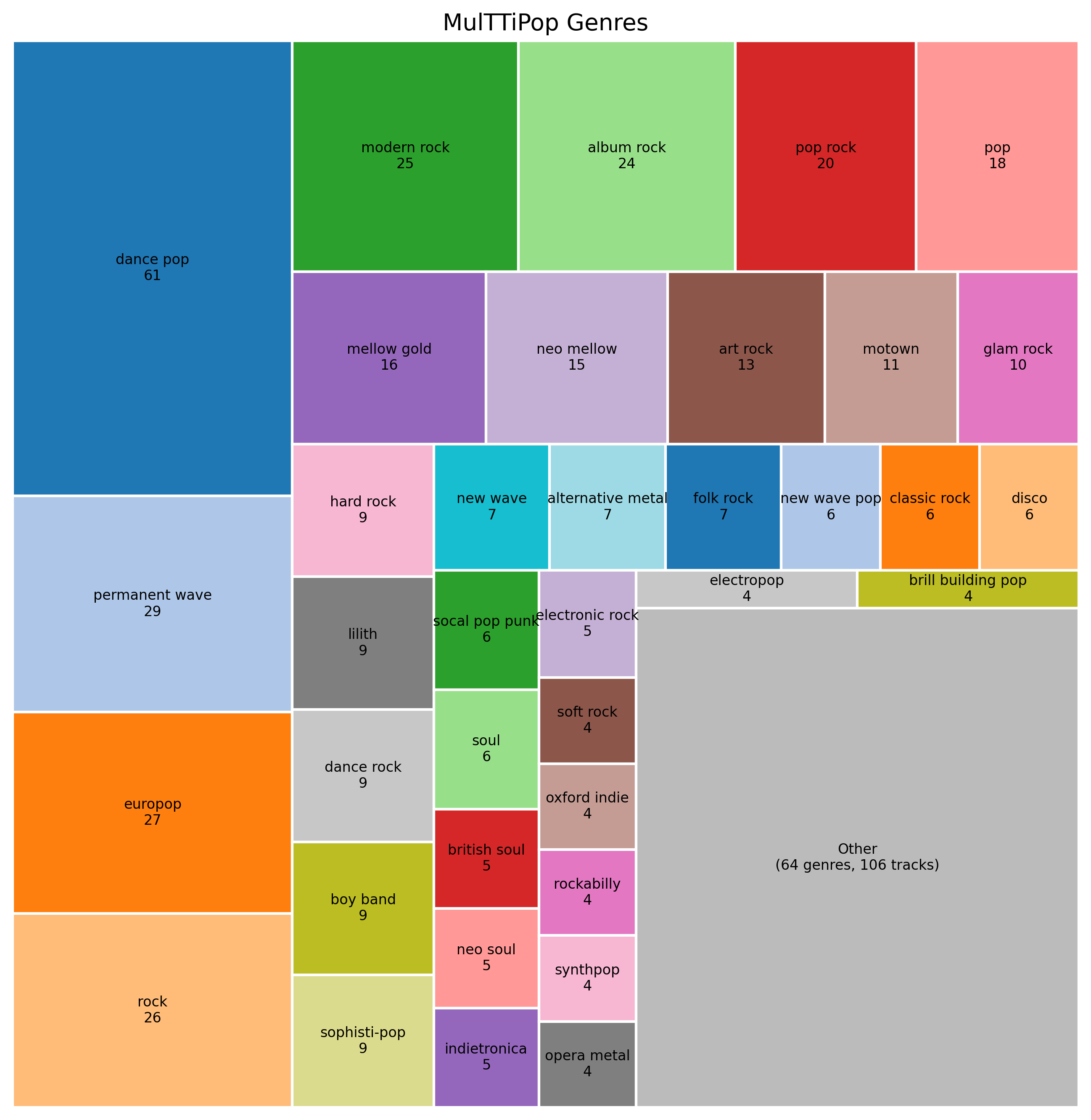}
    \caption{Visualization of genres present in \dsname{}}
    \label{fig:genreviz}
\end{figure*}

To report the number of genres in \dsname{}, we look up the artist of each song using Every Noise at Once \footnote{\url{https://everynoise.com}} and label each song by the first genre displayed. While many songs fall under broad labels such as "dance pop", "rock", and "pop" via this method, some artists fall into specific sub-genres, allowing us to explore the genre diversity of this dataset. We present a treemap of genres present in \dsname{} in Figure \ref{fig:genreviz}.

We calculate the number of genres in RWC via the official description of the dataset \cite{goto2002rwc} \cite{goto2003rwc}:

\begin{itemize}
    \item \textbf{RWC-Pop}: Counted as 2 genres based on the dataset description: "songs on the American hit charts of the 1980s" and "Japanese hit charts of the 1990s".
    \item \textbf{RWC-Jazz}: Counted as 7 genres: "standard-style" jazz, 5 "Style variations", and one "Fusion" genre.
    \item \textbf{RWC-Genre}: Listed as containing 34 sub-genres.
\end{itemize}

We recognize that genre definition and classification is subjective and cannot be reduced to an exact science, and encourage readers to make their own judgment about the genre diversity of the datasets we analyze.


\end{document}